\documentclass{aastex631}
\usepackage{xcolor}
\usepackage{enumitem}
\usepackage[utf8]{inputenc}
\usepackage{textcomp}
\usepackage[utf8]{inputenc}
\usepackage{hyperref}
\DeclareUnicodeCharacter{2212}{-}
\begin{document}
\graphicspath{{./}{Figures/}}


\author[0009-0001-8700-4607]{Ananya Hore}
\email{ananyahore@iisc.ac.in}

\author[0000-0002-4409-7284]{Prantika Bhowmik}
\email{pbhowmik@iisc.ac.in}
\affiliation{
Department of Physics, Indian Institute of Science, Bangalore, Karnataka 560012, India}


\title{\Large Uncovering the Dominant Spatial Scales of the Sun’s Magnetic Field in Solar Cycle 24}

\begin{abstract}
The internal dynamics of the Sun generate magnetic and plasma structures in the photosphere and overlying atmosphere across a wide range of spatial scales. Identifying the critical spatial scale is essential for interpreting physical processes, selecting appropriate observations, optimizing numerical simulations and guiding future instrumentations and space missions. With the growing availability of high-resolution data, we investigate the spatial resolution required to capture the global evolution of the photospheric and atmospheric magnetic field during sunspot cycle 24. We address this problem using a quantitative spherical-harmonic-based modal decomposition. Full-disk photospheric magnetic fields are obtained from the Michelson Doppler Imager and the Helioseismic and Magnetic Imager. The corresponding global coronal field is derived using a newly developed potential field source surface extrapolation (PFSSE) model applied to observed magnetograms. Our analysis reveals two robust results. First, more than $80\%$ of the total modal power is captured by low harmonic degrees corresponding to a spatial scale of approximately 145 Mm, which is significantly larger than individual sunspot dimensions. Second, the effective harmonic degree decreases with height in the corona, indicating that sunspots have little direct and immediate influence on the quasi-static global coronal field. It explains the PFSSE model's success in reproducing and predicting large-scale solar coronal structures, such as streamers and open-field variations, throughout the solar cycle. These results highlight the continued relevance of low-resolution magnetic maps from historical solar observations and contemporary stellar surveys, with direct implications for mission design and for characterizing magnetic environments relevant to exoplanetary systems.\\
\\
\textit{Unified Astronomy Thesaurus concepts:}
\href{http://astrothesaurus.org/uat/1475}{Solar activity (1475)}; 
\href{http://astrothesaurus.org/uat/1503}{Solar magnetic fields (1503)}; 
\href{http://astrothesaurus.org/uat/1492}{Solar evolution (1492)}; 
\href{http://astrothesaurus.org/uat/1527}{Solar surface (1527)}; 
\href{http://astrothesaurus.org/uat/1477}{Solar atmosphere (1477)}; 
\href{http://astrothesaurus.org/uat/1487}{Solar cycle (1487)}; 
\href{http://astrothesaurus.org/uat/1653}{Sunspots (1653)}; 
\href{http://astrothesaurus.org/uat/711}{Heliosphere (711)}

\end{abstract}

\section{Introduction}
Solar phenomena and features span a wide range of spatial and temporal scales, from nanoflares \citep{1988ApJ...330..474P} to heliospheric structures extending $\sim$30–200 Earth radii into the solar wind \citep{DiMatteo2024}. Although the temporal duration of the nanoflares is only a few seconds, heliospheric structures can survive for months. Regardless of their diversity, the majority of these activities are magnetically driven and can be linked to the magnetohydrodynamic dynamo operating in the solar convection zone \citep{2007AIPC..919...49C, Charbonneau2020}. The most prominent and well-studied solar feature (or phenomenon), sunspots, are the regions of intense and concentrated magnetic flux on the solar surface that typically have sizes of $\sim$10$^4$–10$^5$ km and field strengths of several hundred Gauss \citep{mandal2016}. Sunspots are the surface manifestation of the solar dynamo process, and the Babcock–Leighton (B–L) mechanism redistributes the sunspot-associated flux across the photosphere through advection and diffusion driven by large-scale flows and turbulent diffusivity \citep{1961ApJ...133..572B, 1969ApJ...156....1L}. This process produces large-scale diffuse magnetic structures and drives the reversal of the global dipole moment over the solar cycle \citep{Bhowmik2018, Bhowmik_2023}. The solar atmospheric (i.e., coronal) magnetic field responds to changes in the surface field on Alfvénic timescales \citep{2007Sci...317.1192T}. Sunspots (or active regions, ARs) host transient eruptive phenomena associated with the solar corona, including flares and coronal mass ejections (CMEs). Excluding such transients, the global corona evolves quasi-statically and remains closely coupled to the large-scale surface magnetic field produced by flux transport processes \citep{Mackay_2012}, resulting in a much slower evolution of large-scale coronal structures such as streamers and coronal holes.

These considerations suggest that global simulations of the solar surface and corona must, at a minimum, resolve length scales associated with sunspots. The number of sunspots emerging on the surface varies over the 11 yr solar cycle, with solar minima dominated by weak, large-scale fields and maxima characterized by numerous sunspots concentrated in activity belts \citep{Hathaway2010, articleIva}. Consequently, for global surface simulations, lower spatial resolution should suffice during minima, whereas higher resolution should be required to capture sunspot-associated magnetic field distribution during maxima. Quasi-static global coronal field simulations are expected to exhibit similar resolution requirements.

Progress in global magnetic field modeling has been tightly coupled to advances in observational resolution. Early full-disk magnetograms from the Wilcox Solar Observatory (WSO; $\sim$$3'$) were unable to resolve sunspots \citep{1977SoPh...55...63D}. Subsequent space-based observations significantly improved spatial resolution, including the Michelson Doppler Imager (MDI) on board the Solar and Heliospheric Observatory (SOHO; $\sim$$4''$; \citeauthor{Meunier_1999} \citeyear{Meunier_1999}) and, more recently, the Solar Dynamics Observatory (SDO) Helioseismic and Magnetic Imager (HMI) and Atmospheric Imaging Assembly (AIA). HMI provides full-disk vector magnetograms at $\sim$$1''$ resolution ($\sim$725 km; \citeauthor{Scherrer2012} \citeyear{Scherrer2012}), while AIA images the corona with an effective resolution of $\sim$$1''.5$ ($\sim$1100 km; \citeauthor{2012SoPh..275...17L} \citeyear{2012SoPh..275...17L}). AR-focused Space-weather HMI Active Region Patch (SHARP) data achieve spatial scales of $\sim$365 km \citep{Bobra2014}. At still finer scales, the ground-based Daniel K. Inouye Solar Telescope (DKIST) now resolves magnetic and plasma structures down to $\sim$20–30 km in localized regions \citep{article, Kuridze_2025}.

Assimilating such ultra-high-resolution observations into global simulations with uniform grid spacing is computationally expensive. This raises a fundamental question: What is the maximum spatial resolution required to capture the global evolution of the Sun’s magnetic field? We address this question for solar cycle 24 (2009–2020) using a mathematical and quantitative approach based on spherical harmonic decomposition. Full-disk MDI and HMI magnetograms are used to characterize the surface field. For the corona, where global magnetic field measurements are unavailable, we employ potential field source surface extrapolation (PFSSE; \citeauthor{1969SoPh....9..131A} \citeyear{1969SoPh....9..131A}; \citeauthor{1969SoPh....6..442S} \citeyear{1969SoPh....6..442S}). In the low-$\beta$ corona, quasi-static equilibrium requires a vanishing Lorentz force, making PFSSE a suitable first-order approximation for large-scale structures \citep{2009SoPh..260..321M, Mackay_2012, 2019SpWea..17.1293N}. Modal analysis of both surface and coronal fields provides a quantitative measure of the dominant length-scales throughout solar cycle 24. The universality of the outcome from our analyses suggests that the conclusions should also be valid for other solar cycles. It is important to note that because of the current-free assumption, PFSSE does not account for nonpotential magnetic structures in the corona. Studies have demonstrated that nonpotential magnetic fields associated with large and midscale ARs can introduce distortions in coronal field configuration, leading to differences compared to potential-field extrapolation \citep{Riley_2006}. Full and realistic MHD models could provide better insights into these effects. However, exploring such nonpotential contributions is beyond the scope of the present study, and we restrict our analyses to the potential-field framework.

This study is structured as follows. Section \ref{sec:obsm} describes the observational data, and Section \ref{sec:method} outlines the methodology of modal decomposition, associated analyses, and a brief of our newly developed coronal model. Section \ref{sec:result} presents the results corresponding to both observed surface and modeled coronal magnetic field distributions, and Section \ref{sec:conclusion} concludes with summarizing the implications and applicability of our findings.

\section{Observational Data} \label{sec:obsm}
Our analysis of the relevant spatial scales combines surface magnetic field measurements with coronal extrapolation, beginning with observed photospheric magnetograms. We use full-disk synoptic maps of the radial magnetic field ($B_r$) from SOHO/MDI and SDO/HMI, including polar field corrections, covering Carrington rotations from 2009 to 2020. In total, 158 Carrington maps are analyzed. The original MDI ($3600\times1080$) and HMI ($3600\times1440$) maps are downsampled to a common grid of $361\times181$ points in longitude and latitude. To validate the simulated coronal magnetic structures, we additionally compare our results with 10 minute averaged K-Cor coronagraph observations from the Mauna Loa Solar Observatory \citep{kcor2013}.

\section{Methodology} \label{sec:method}

Details of the modal decomposition technique and the coronal magnetic field model used in this study are provided in the appendices. Here we summarize the approach. Characteristic spatial scales in the photosphere and corona are quantified using spherical harmonic decomposition. The observed surface magnetic field distribution (radial component, $B_{R_\odot}(\theta, \phi)$) is expressed using spherical harmonics ($Y_l^m$) in the spectral space as 
\begin{equation}
B_{R_\odot}(\theta, \phi)= \sum_{l=1}^{\infty}\,\sum_{m=-l}^{l} C_{l}^{m}\, Y_l^m (\theta, \phi)
\label{eq:modemag}
\end{equation}
where, $C_{l}^{m}$ is the modal contribution corresponding to degree $l$ and order $m$, while $\theta$ and $\phi$ denote colatitude and longitude, respectively (for more details, see Appendix \ref{Appen:A}). The upper limit of $l$ can be varied depending on the required resolution. Moreover, the mode contribution coefficient is calculated as follows \citep{stansby2022testproblemspotentialfield}:

\begin{equation}
    C_{l}^{m} = \int_{\phi = 0}^{2\pi} d\phi\int _{\theta =0}^{\pi}sin\,\theta\, d\theta \, B_{R_\odot}(\theta,\phi)\, Y_l^{m*}(\theta,\phi)
    \label{eq:coeff}
\end{equation}
Now, $B_{R_\odot}(\theta,\phi)$, representing the observed surface magnetic field distribution, can be replaced by any other 2D distribution of magnetic fields [$B_{r}(\theta,\phi)$] on the surface of a sphere placed at different heights (or radius, $r$) in the corona. The corresponding $C_l^m$ will provide the spatial scale particulars at those heights. The magnetic field distributions in the quasi-static solar corona, extended from the surface to $2.5 \, R_\odot$, are evaluated using our newly developed PFSSE model (for more details, see Appendix \ref{Appen:B}). The bottom boundary of the PFSSE model is the observed magnetogram maps.

We further compute various quantities to provide a comprehensive quantitative analysis of the observed photospheric and simulated coronal magnetic field distribution during solar cycle 24. This allows us to investigate a range of scientific questions concerning large-scale solar magnetic topology and distributions.

\textit{Effective mode contribution}. Modal coefficients with different $l$ and $m$ values are calculated for any given 2D magnetic field maps (either on the surface or at a specific coronal height). Using them, we define the effective contribution normalized over order $m$ for each $l$ given by the following expression:
\begin{equation}
C_l = \frac{\Sigma_m |C_{l}^{m}|}{\Sigma_m |m|}
\label{eq:effmode}
\end{equation}

\noindent which indicates the relative importance of different spherical harmonic degrees. Degree $l$ corresponds to an approximate spatial scale of $\frac{2\pi R_\odot}{\sqrt{l(l+1)}}$ \citep{Luo_2023}. The effective distribution of $C_l$ for individual Carrington rotations provides a quantitative estimate of the maximum-$l$ required to capture global magnetic structures both in the surface and the overlying corona at any given time during solar cycle 24.

\textit{Total unsigned magnetic flux}. We compute the global radial magnetic flux at different heights \citep{2019SpWea..17.1293N},
\begin{equation}
\Phi(r) = r^2 \int_{0}^{2\pi} d\phi \int_{0}^{\pi} |B_r(r,\theta,\phi)|\sin\theta\, d\theta,
\label{eq:flux}
\end{equation}
as an additional measure of scale-dependent evolution.

\textit{Tracing of magnetic field lines}. Field lines are traced to the source surface ($2.5 \, R_\odot$) for visualization and as a complementary diagnostic of the spatial resolution required in global coronal simulations.

\section{Results} \label{sec:result}
This section presents two analyses that yield striking conclusions. The first one focuses on applying modal decomposition to both observed surface and simulated coronal magnetic field distributions during solar cycle 24. The second part highlights the role of emerging sunspots in quasi-static coronal magnetic field structures, with a complementary comparison to coronal white-light observations.

\subsection{Mode Analysis and Relevant Spatial Scales from the Photosphere to the Corona}
\label{subsec:result_mode_analysis}
\begin{figure}
\begin{minipage}[t]{0.34\textwidth}
  \centering
  \includegraphics[width=\linewidth]{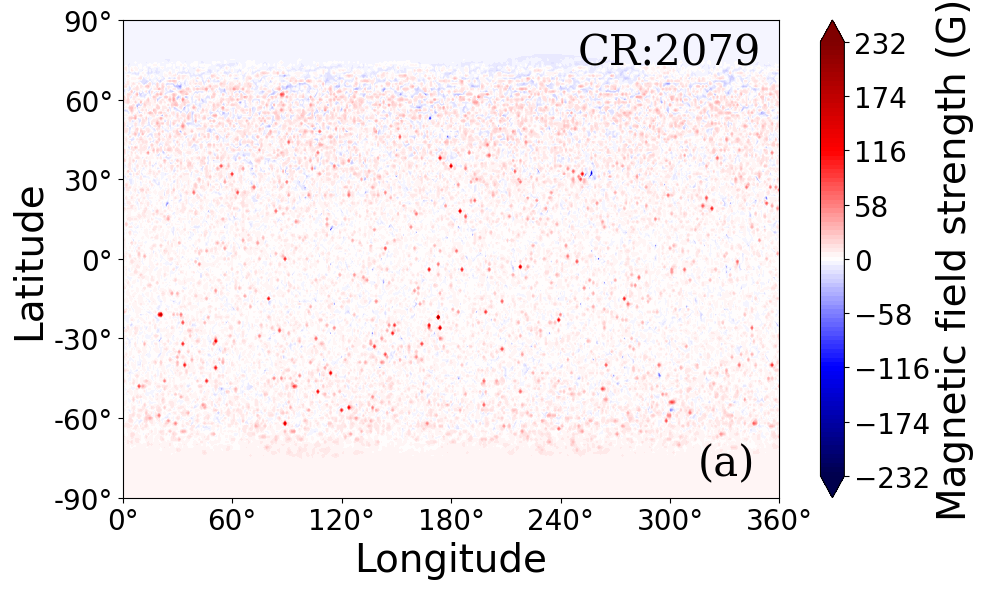}
\end{minipage}
\hfill
\begin{minipage}[t]{0.34\textwidth}
  \centering
  \includegraphics[width=\linewidth]{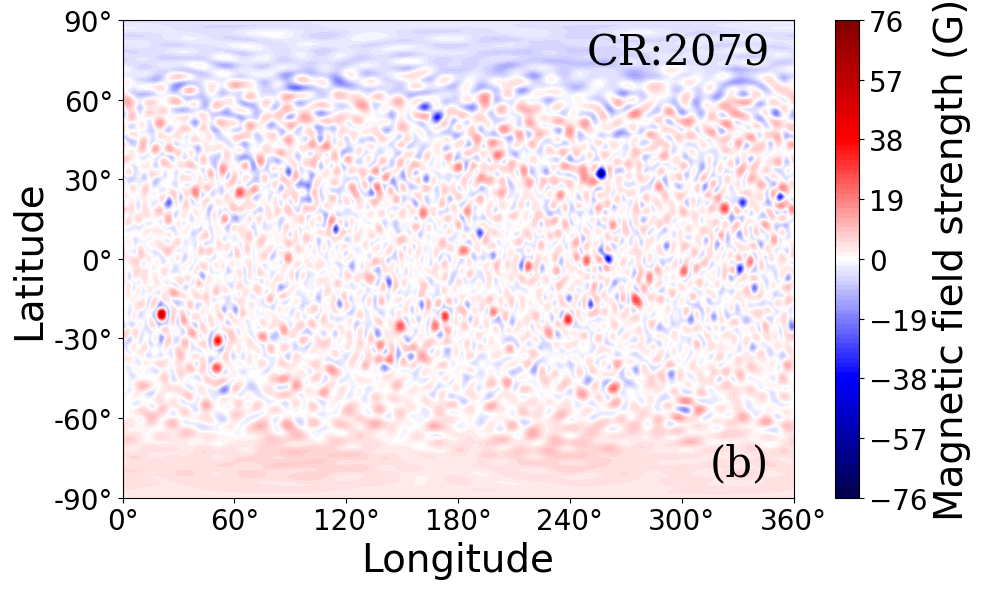}
\end{minipage}
\hfill
\begin{minipage}[t]{0.27\textwidth}
  \centering
  \includegraphics[width=\linewidth]{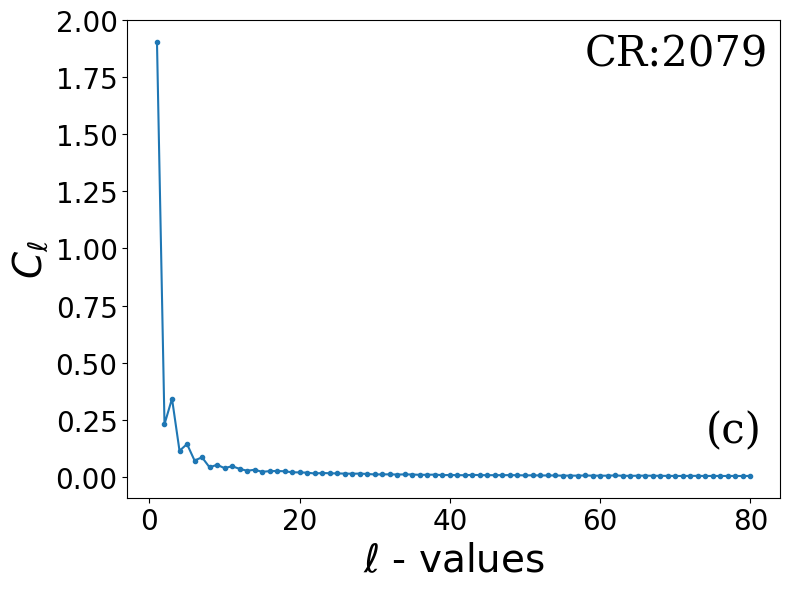}
\end{minipage}

\begin{minipage}[t]{0.34\textwidth}
  \centering
  \includegraphics[width=\linewidth]{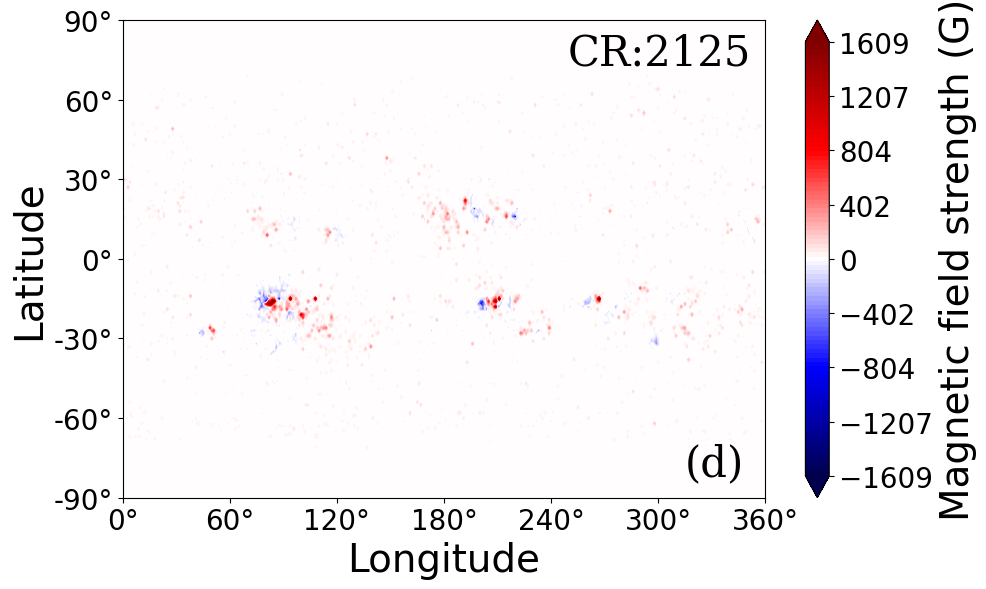}
\end{minipage}
\hfill
\begin{minipage}[t]{0.34\textwidth}
  \centering
  \includegraphics[width=\linewidth]{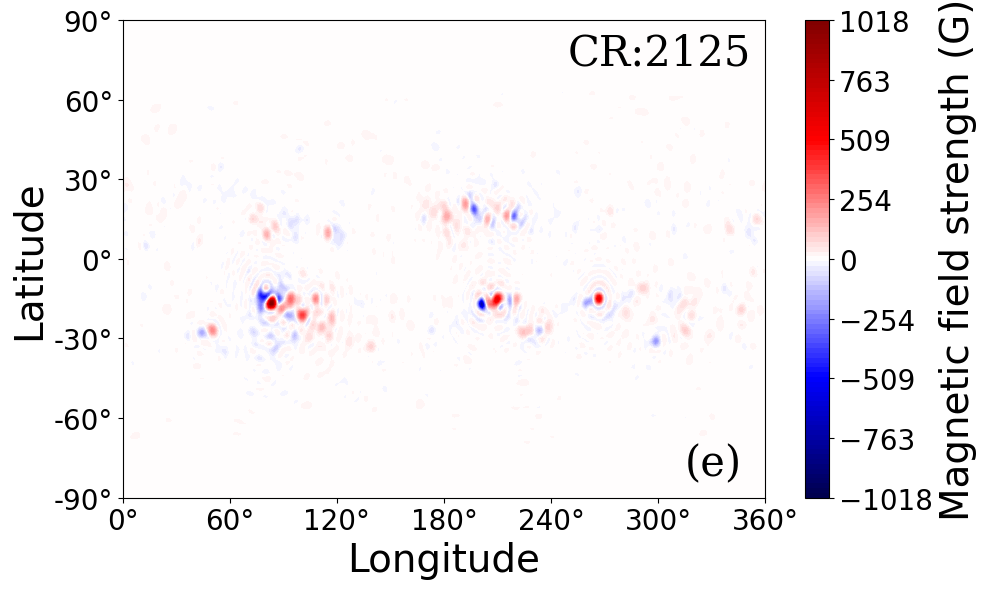}
\end{minipage}
\hfill
\begin{minipage}[t]{0.27\textwidth}
  \centering
  \includegraphics[width=\linewidth]{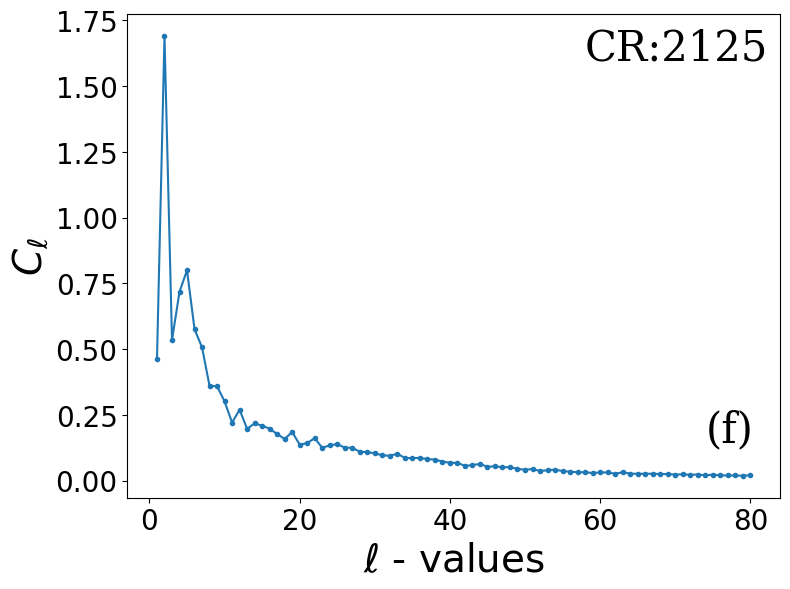}
\end{minipage}
\caption{Surface magnetic maps and effective modes for two representative cases of Carrington rotations 2079 and 2125. For CR 2079, (a) the observed synoptic magnetogram saturated at $\pm \,232$ G, (b) the reconstructed photospheric magnetic map with $l_\mathrm{max}=80$, and (c) the effective mode contribution are presented in the top row. Similarly, for CR 2125, (d) the observed synoptic magnetogram saturated at $\pm \, 1609$ G, (e) the reconstructed photospheric magnetic map with $l_\mathrm{max}=80$, and (f) the effective mode contribution are in the bottom row. The color bars on the right side of the magnetic maps represent the field strength in each case.}
\label{fig:2079_2125}
\end{figure}

We begin with two representative Carrington rotations of the observed photospheric radial field ($B_r$): solar minimum in 2009 (CR 2079) and the magnetic dipole reversal phase in 2012 (CR 2125, near cycle-24 maximum). For both cases, the observed maps are reconstructed using spherical harmonics truncated at $l_{\rm max}=80$, and the effective mode contributions (Equation (\ref{eq:effmode})) are evaluated. Figures \ref{fig:2079_2125}(a) and (b) show the observed and reconstructed (using $l_{\rm max}=80$) maps for CR 2079. Increasing $l_{\rm max}$ progressively resolves smaller-scale structure until the observational limit is reached, but the required $l_{\rm max}$ depends on whether one aims to fully reproduce the observed magnetogram to the smallest scale or to retain only the dominant contributing modes. Figure \ref{fig:2079_2125}(c) demonstrates the effective mode contribution spectrum for individual $l$ calculated using Equation (\ref{eq:effmode}) during solar minimum.

It is evident that during CR 2079, the magnetic topology is dominated by large-scale modes. The effective spectrum peaks at $l=1$ (dipole), followed by $l=3$, with rapidly declining contributions at higher $l$. Increasing $l_{\rm max}$ adds progressively finer structures, approaching the intrinsic resolution of the magnetograms. The solar minimum field distributions have strong polar concentrations with opposite signs in the two hemispheres, reflecting the dominance of odd-$l$ parity. The corresponding similar analysis associated with CR 2125 (close to solar cycle 24 maximum) is presented in Figures \ref{fig:2079_2125}(d), (e), and (f). The strongest contribution arises from $l=2$ (quadrupole), followed by $l=4$ and $5$, while contributions for $l>5$ become negligible. Near the epoch of solar maximum and dipole reversal, the prevalence of even-$l$ modes corresponds to weak polar fields and the quadrupolar distribution suggests the sunspot activity belts at low latitudes. The shift from odd-$l$ dominance in minima to even-$l$ dominance during reversal explains the contrasting large-scale topologies.

\begin{figure}
  \centering
  \begin{minipage}{0.9\textwidth}
    \includegraphics[width=\linewidth]{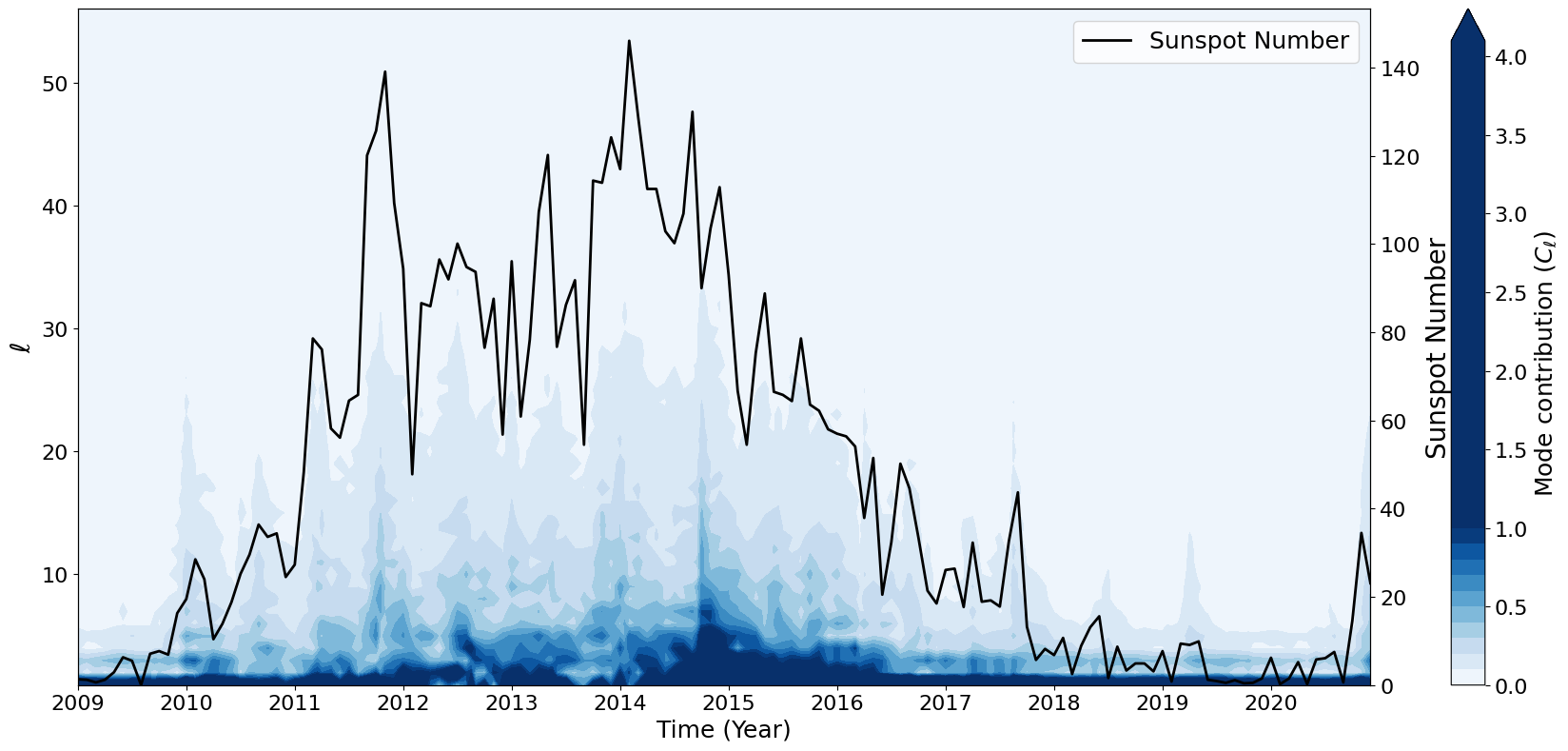}
\caption{Effective mode contribution spectrum on the photosphere for solar cycle 24. The color bar is saturated at a maximum mode strength of $C_l = 1.0$. The black curve represents the sunspot cycle 24 with average sunspot numbers corresponding to each Carrington rotation.}
    \label{fig:con}
    \end{minipage}
    \end{figure}

\begin{figure}[htbp]
  \centering
  \begin{minipage}{0.9\textwidth}
  \centering
    \includegraphics[width=\linewidth]{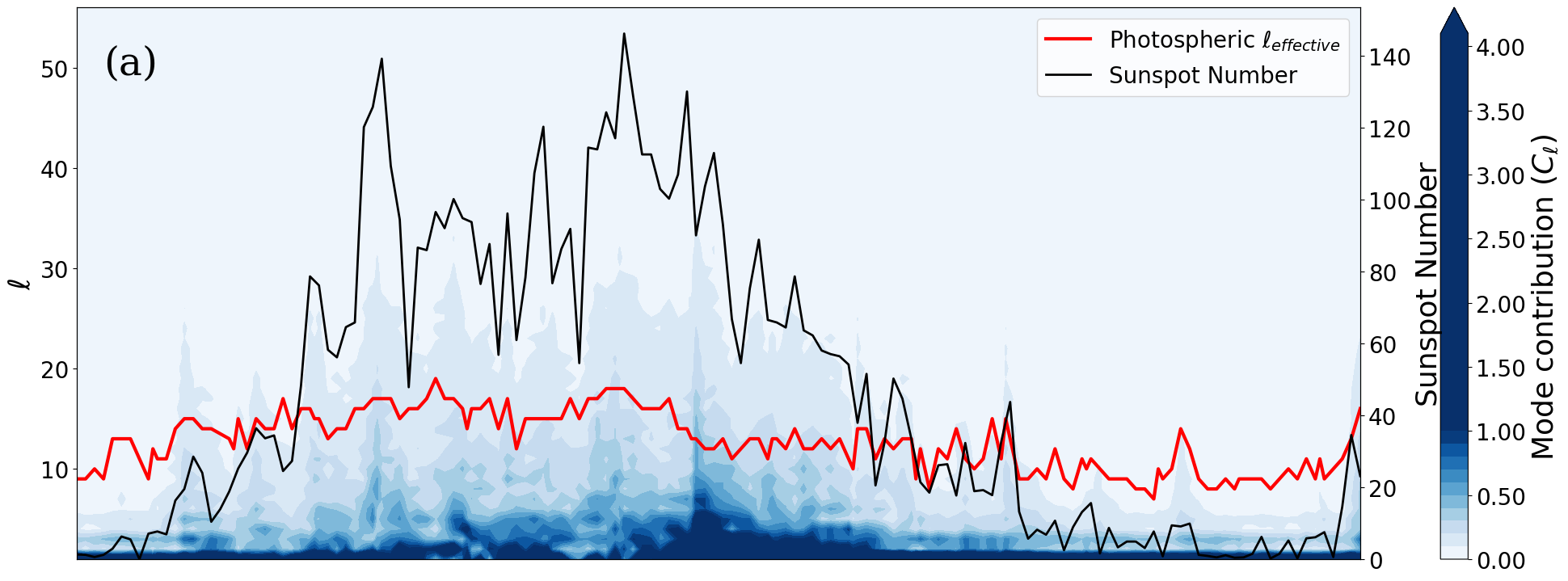}
    \end{minipage}

 \begin{minipage}{0.9\textwidth}
  \centering
    \includegraphics[width=\linewidth]{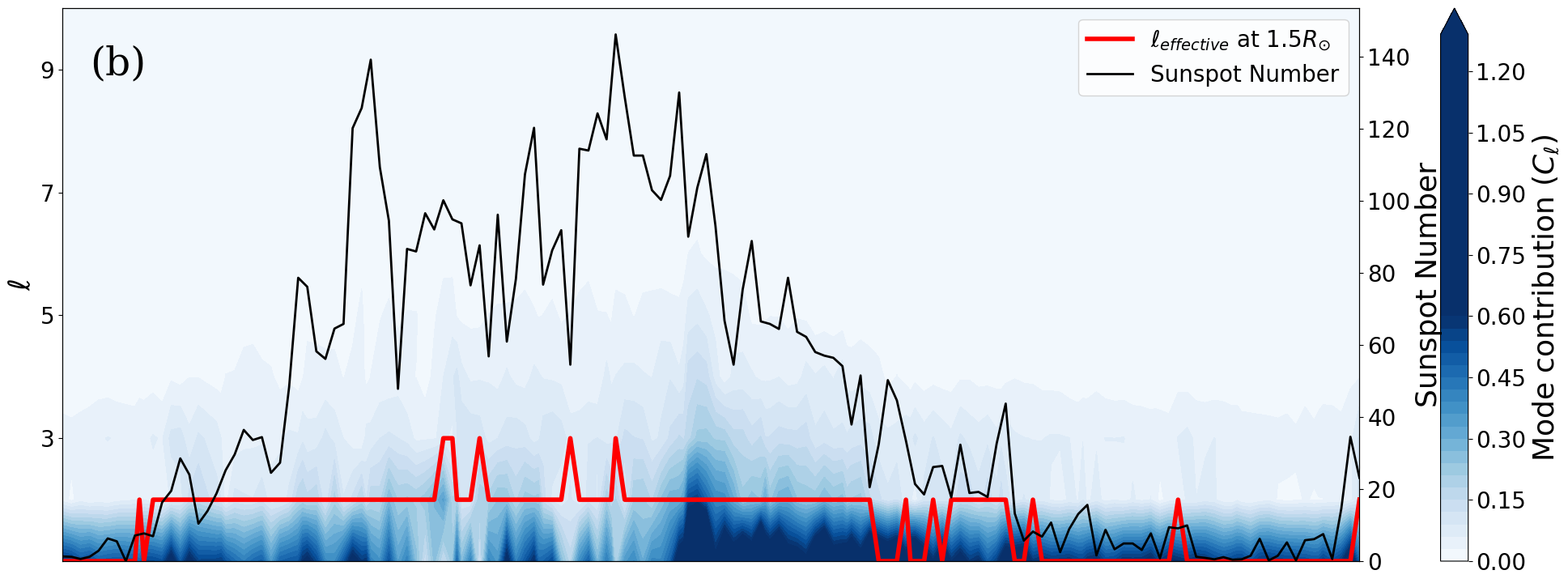}
    \end{minipage}

\begin{minipage}{0.9\textwidth}
  \centering
    \includegraphics[width=\linewidth]{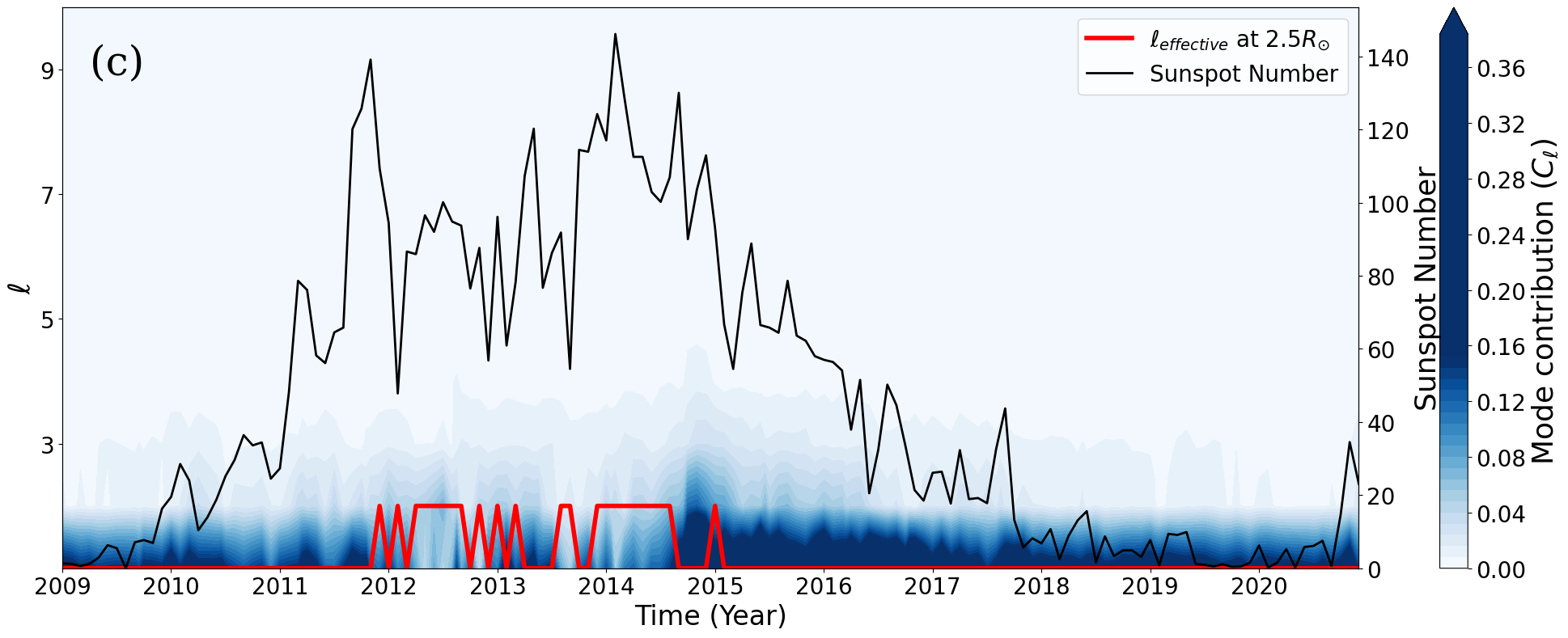}
    \end{minipage}
  \caption{Effective mode contribution spectra for (a) the photosphere and (b), (c) two different heights inside the corona: (b) $1.5\, R_{\odot}$ and (c) $2.5\, R_{\odot}$, superimposed with sunspot number cycle 24 (black curve) and $l_\mathrm{effective}$ for that corresponding height (red curve). The color bars are saturated at the maximum mode strength of (a) $C_l=1.00$ associated with the solar photosphere, (b) $C_l=0.60$ and (c) $C_l=0.15$ associated with coronal heights of $1.5\, R_{\odot}$ and  $2.5\, R_{\odot}$, correspondingly.}
  \label{fig:coronaspc}
\end{figure}

We extend the effective mode contribution analysis to all Carrington rotations of solar cycle 24. The result, shown as a contour map overlaid with the sunspot number (Figure \ref{fig:con}), reveals two key features. First, the magnetic power is strongly concentrated at low spherical harmonic degrees, with dominant contributions from $l<10$ and rapidly diminishing power at higher $l$. Although the analysis is performed up to $l_{\rm max}=80$ ($\sim$55 Mm), the spectrum indicates that $l_{\rm max}\approx30$ ($\sim$145 Mm) is sufficient to capture the dominant surface magnetic structures throughout the cycle. In fact, it reaches a maximum value of $l_{\rm max} = 34$ ($\sim$128 Mm) during Carrington rotation 2156 ($2014$ October 14 − $2014$ November 11). Second, this low-degree dominance persists even during solar maximum, despite the increased sunspot number. While higher $l$ modes progressively resolve smaller-scale features, the global surface magnetic field remains governed by large-scale modes at all phases of the activity cycle. We further find that $80\%$ of the total spectral-coefficient (area under the curve of the spectral-mode contributions) for most of the Carrington maps during solar cycle 24 is limited to $l^* = 30$, while it decreases further to $l^*=18$ at solar minimum (CR 2079). Details on the associated analyses are given in Appendix \ref{Appen:C}. To quantify the dominant effective harmonic degree, we further compute the following quantity:
\begin{equation}
l_\mathrm{effective} = \frac{\sum_l l\,C_l}{\sum_l C_l},
\label{eq:effl}
\end{equation}
which varies between $l_\mathrm{effective}\approx7$ and $19$ over the cycle 24 (see, Figure \ref{fig:coronaspc}(a)) for the surface magnetic field. For all Carrington rotation maps, using $l_\mathrm{max}\approx l_\mathrm{effective}$ captures more than $60\%$ of the total contribution, corresponding to characteristic spatial scales of $\sim$226–588 Mm. Notably, $l_\mathrm{effective}$ decreases further between mid-2014 to early 2015, indicating that the dominant power is concentrated in lower harmonics and that high-$l$ modes are unnecessary to recover most of the global surface magnetic structures close to the solar cycle maximum.

We next focus on coronal magnetic field structures and the required resolution associated with the relevant spatial scale. We utilize the PFSSE model discussed in Appendix \ref{Appen:B} to extrapolate the coronal field from each Carrington map. To find out how the different harmonic modes' contributions vary as a function of coronal altitude, we select two specific heights: $1.5\, R_{\odot}$ and $2.5\, R_{\odot}$ (the source surface). The resultant contour maps for the mode contributions for both heights are shown in Figures \ref{fig:coronaspc}(b)-(c). Similar to the photospheric pattern, we notice that the mode contributions ($C_l$) diminish very sharply with increasing $l$ for both coronal heights. Likewise, the clustering of harmonic power to very low values of $l$, reduces the $l_\mathrm{effective}$ for all heights in the corona, which is already drastically smaller than the photospheric values (see the red curves in the subplots of Figure \ref{fig:coronaspc}). This emphasizes that the immediate effect of individual emerging sunspots on the quasi-static coronal features reaching heights beyond $1.5\, R_{\odot}$ remains negligible throughout the solar cycle. However, this inference may not be universally applicable. The extent of influence also depends on the strength and magnetic complexity of the sunspot. In principle, a sufficiently strong AR possessing significant twist and nonzero magnetic helicity can modify or even dominate the ambient quasi-static large-scale coronal magnetic field at source surface height. The collective role of sunspots becomes important in the global corona only when they become a part of the large-scale surface features through the surface evolution. A closer inspection of Figure \ref{fig:coronaspc} reveals that, $l_\mathrm{effective}$ pertaining to $1.5\, R_{\odot}$ ranges from $1$ to $3$, while that for $2.5\, R_{\odot}$ is predominantly dipolar with $l_\mathrm{effective}=1$ (except for a few Carrington rotations when the effective degree is quadrupolar with $l_\mathrm{effective}=2$). Thus, beyond a distance ($\sim 1.5 \,R_\odot$ from the center of the Sun), the field predominantly exhibits dipolar, quadrupolar, or octopolar characteristics, in accordance with the solar cycle phase. 

\subsection{Role of Emerging Sunspot in Quasi-static Corona}
\label{subsec:result_corona}
\begin{figure}[htbp]
  \centering
  \begin{minipage}{0.9\textwidth}
  \centering
    \includegraphics[width=\linewidth]{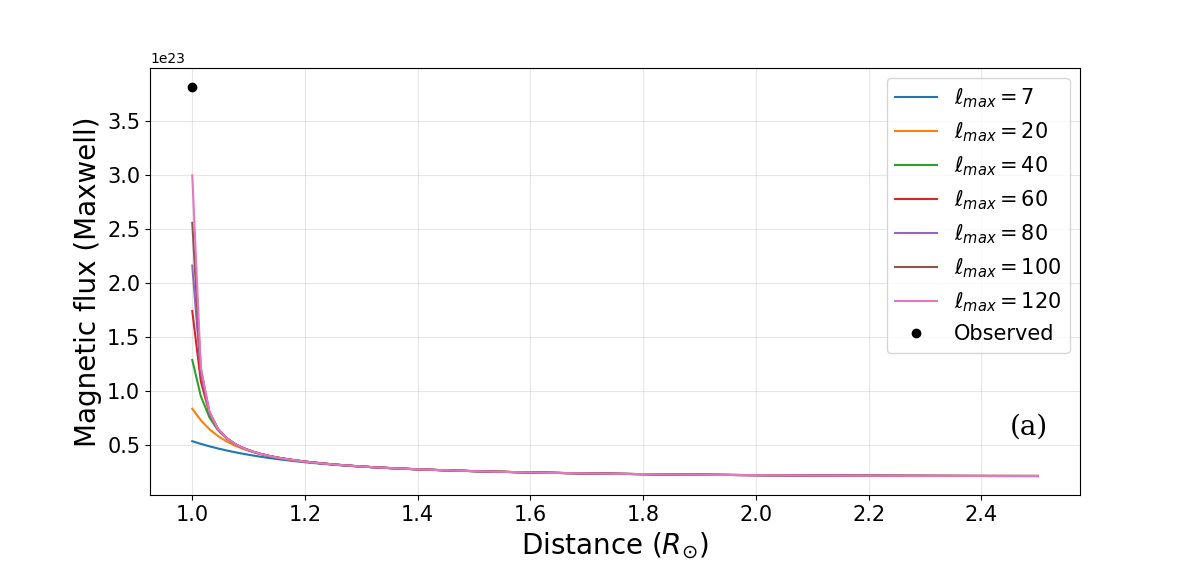}
    \end{minipage}
    \begin{minipage}[t]{0.32\textwidth}
  \centering
  \includegraphics[width=\linewidth]{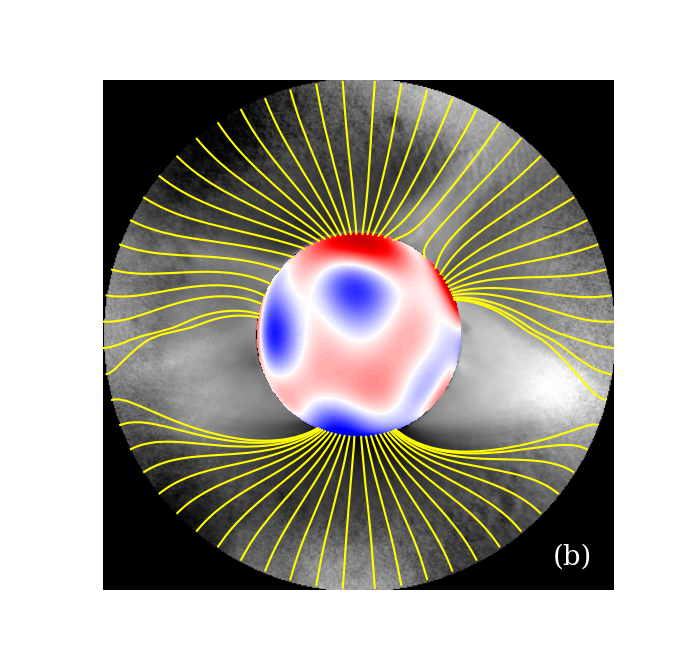}
\end{minipage}
\hfill
\begin{minipage}[t]{0.32\textwidth}
  \centering
  \includegraphics[width=\linewidth]{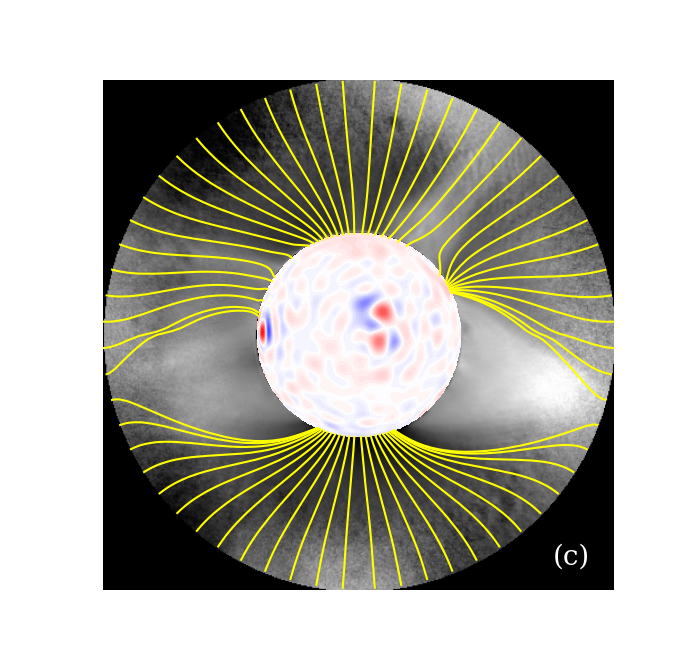}
\end{minipage}
\hfill
\begin{minipage}[t]{0.32\textwidth}
  \centering
  \includegraphics[width=\linewidth]{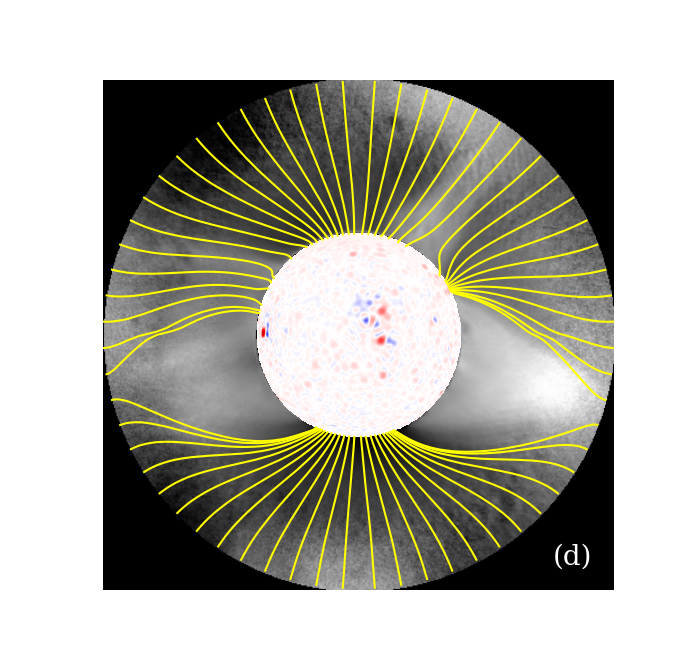}
\end{minipage}
\caption{Magnetic flux distribution and coronal field lines illustrate the diminishing influence of individual sunspot with increasing height. (a) Flux variation with distance from photosphere for Carrington rotation 2081. Observed photospheric flux and fluxes for different values of $l_\mathrm{max}$ are denoted with different colors. (b)-(d) Open magnetic field lines superimposed on K-Cor coronagraph image acquired on 2017 July 25. Field lines from the photosphere are reaching the source surface at $2.5\, R_{\odot}$ for (b) $l_\mathrm{max}=7$, (c) $l_\mathrm{max}=30$ and (d) $l_\mathrm{max}=80$ surface maps. In panels (b)-(d), solar north is up.}
\label{fig:kcorline}
\end{figure}

Using the derived coronal field, we carry out an assessment of flux variation with altitude inside the corona. Figure \ref{fig:kcorline}(a) demonstrates this variation of total unsigned magnetic flux with upward radial distance from the photosphere for the Carrington rotation 2081 ($2009$ March 9 − $2009$ April 5). In the photosphere, the fluxes corresponding to different values of $l_\mathrm{max}$ deviate largely from each other and also from the one calculated using the observed field distribution. However, they all attenuate very fast to some constant value within a height $\sim 1.15\, R_{\odot}$, roughly corresponding to the upper limit of the transition region. Beyond that height, the differences gradually decrease and become negligible within the corona. Thus, it suggests that even a very low $l_\mathrm{max} \approx 5-10$ is enough to provide a rough estimate of the characteristic spatial scale capturing the key magnetic features of the global solar corona. It is consistent with the fact that WSO uses $l_\mathrm{max}=9$ to construct magnetic field maps at the source surface \citep{1984PhDT.........5H, refId0}. Similarly, previous studies have also shown that the mean coronal flux density saturates beyond $l_\mathrm{max}=4$ for heights $\ge2.0\,R_{\odot}$, while values up to $l_\mathrm{max}=9$ are required at lower heights ($\sim 1.5\, R_{\odot}$; \citeauthor{2019A&A...626A..67V} \citeyear{2019A&A...626A..67V}). This further supports our result regarding the adequacy of low $l_\mathrm{max}$. In addition, it indicates that the effect of individual sunspots continues to diminish with increasing altitude. The mathematical formulation, which dictates how the components of the magnetic field vary as $\sim (\frac{R_{\odot}}{r})^{l+2}$ with height, also justifies our finding. So, the effects of the higher harmonics associated with the individual sunspots fade with increasing height inside the atmosphere, leaving only the large-scale structures surviving. However, we caution readers that the same conclusion cannot be drawn for analyses based on the magnetic flux associated with the surface magnetic field distribution. For example, during a quiet phase of cycle (solar minimum), a very few emerged sunspots will require high $l_\mathrm{max}$ to achieve $60\,\%$ of magnetic flux coverage (for more details, refer to Appendix \ref{Appen:D}).

In the context of solar corona, we continue to test the same conjecture by comparing the simulated and observed corona for a particular Carrington Map (CR 2193) corresponding to the American solar eclipse of 2017
\citep{2018ApJ...853...72N}. We compare our model extrapolated coronal field structures based on three different lower boundary surface maps corresponding to $l_\mathrm{max}=7$, $l_\mathrm{max}=30$, and $l_\mathrm{max}=80$, respectively, superimposed on the 10 minute average K-Cor coronagraph image from MLSO for 2017 July 25 \citep{kcor2013}. Figures \ref{fig:kcorline}(b)-(d) represent extrapolated open magnetic field lines against the observed corona where all three of the simulated coronae matched the observed large-scale structures like helmet streamers, pseudo-streamers, etc. Thus, it demonstrates the PFSSE model's ability to accurately capture the bright, quasi-static, large-scale structures extending outward, even with a large deviation in surface field resolution. Moreover, although the coronal field distribution is a little bit different near the photosphere for $l_\mathrm{max}=7$, $l_\mathrm{max}=30$, and $l_\mathrm{max}=80$, these deviations are negligibly small and disappear with height. This highlights the effectiveness of low-resolution ($l_\mathrm{max}\approx 7$) surface maps utilized for capturing global coronal structures, excluding transients such as CMEs. 

\section{Concluding discussions} \label{sec:conclusion}

Spatial and temporal resolutions are crucial for any computational model that mimics physical phenomena. The required or implemented resolution can be dictated by the resolution of the observational data to be compared with or assimilated into the simulations, computational resource limitations, or both. We investigate what spatial resolution is relevant to the global photospheric and coronal magnetic fields of the Sun during sunspot cycle 24 (2009-2020). Modal decomposition based on spherical harmonics plays the dominant role in our analyses, which are applied to all 158 Carrington maps associated with observed full-Sun photospheric magnetograms and corresponding PFSSE coronal maps during cycle 24. In our presented work, we provide a quantitative estimate of the \textit{maximum} required spatial resolution by using various diagnostic parameters, including the effective mode contribution ($C_l$), the effective degree ($l_\mathrm{effective}$), and the total unsigned magnetic flux ($\Phi(r)$). 

A lower spherical harmonic degree, $l<7$, corresponds to large-scale structures; for example, $l = 1$ and $2$ correspond to the Sun's global magnetic dipole and quadrupole, respectively. A higher value of $l$, like $l \approx 60$, corresponds to typical length-scales of sunspot radii. Our study demonstrates that, to encompass the major characteristics in the observed full-Sun photospheric magnetic field distribution throughout different phases of cycle 24, using spectral decompositions with even a low $l_\mathrm{max} \approx 20-30$ (corresponding to a minimum spatial scale $\approx 145\, \mathrm{Mm}$) is sufficient. Moreover, the majority of the spectral power remains concentrated within the lower modes, resulting in a further, smaller value for $l_\mathrm{effective} \approx 7-19$ corresponding to a spatial scale $\approx 226\, \mathrm{Mm}$. In particular, the effect of multiple new sunspot emergences during solar cycle maximum collectively contributes to large-scale modes, which contradicts our usual reasoning that computational models aiming to mimic global photospheric magnetic field evolution should have spatial resolutions substantially higher than sunspot radii. 

The photospheric magnetic field distribution primarily dictates the quasi-static coronal magnetic structures. Performing PFSSEs on the 158 Carrington maps, we evaluated the corresponding coronal magnetic field distributions up to $2.5$ $R_\odot$ during cycle 24. Irrespective of the resolution of the surface field, the magnetic flux decreases rapidly with coronal height, and beyond $\sim 1.15\, R_{\odot}$ (equivalent to the upper limit of the transition region), the majority of surviving coronal structures correspond to low-mode harmonics of the surface field. Thus, all PFSSE coronal maps based on three very different values of $l_\mathrm{max}:$ 7, 30, and 80, identically matched the observed white-light corona in our case study. The diminishing effect of individual sunspots on the quasi-static corona with increasing height results in low-mode dominance in coronal magnetic field distributions throughout the solar cycle. For instance, at the source surface ($2.5 R_{\odot}$), the $l_\mathrm{effective}$ alternates between $1$ (dipole) or $2$ (quadrupole) depending on the phase of the sunspot cycle (minima versus maxima).

We conclude with three major highlights and their relevance in diverse aspects. First, the dominance of low spectral modes in the solar surface magnetic field will allow the use of earlier low-resolution observations in global simulations (for example, surface flux transport simulations), even though the current trend emphasizes assimilating high-resolution data. Performing such global simulations will also be convenient, with much lower computational time and resource requirements. Second, the persistent presence of only low modes in the global quasi-static corona indicates that the formation of large-scale coronal magnetic structures (such as streamers) reaching higher altitudes is a slow process, and that the emergence of sunspots has no immediate effect. If sunspots are not associated with any transient phenomena such as CMEs, it may take months before their impact alters magnetic field connectivities at higher heights of the solar corona. Thus, it is possible to predict the global corona during an intended epoch and estimate the Sun's open magnetic flux days or months in advance (depending on the solar cycle phase) - both of which are important in space weather studies and space mission planning. Third, our study indicates how the available observed low-resolution measurements of the stellar surface magnetic field can be utilized to assess the stellar environment, including large-scale stellar coronal magnetic field structures, their extension, and effects on the hosted exoplanets \citep{Reiners2012, 2020pase.conf...89K}. Thus, predicting the ambient space weather of exoplanets would be possible even with low-resolution stellar data \citep{strickert_2024}. We plan to explore these three distinct aspects in our future studies.

\section{Acknowledgment}
The work was carried out at the Indian Institute of Science, Bangalore. A.H. acknowledges the financial support from the fellowship agency, the Ministry of Education (MoE). P.B. acknowledges the support from the Start-up Grant provided by the Institute. For analyses and visualization, open source Python packages have been used \citep{harris2020array, 2020SciPy-NMeth, astropy:2022}. We also appreciate the use of data from NASA’s SDO/HMI, SOHO/MDI, and MLSO/K-Cor and thank their science teams for their efforts in producing and maintaining the observations. Finally, we thank the reviewer for the careful reading of the manuscript and the valuable suggestions.

\bibliographystyle{aasjournal}
\bibliography{references}

\appendix

\section{Spherical Harmonics-based Modal Decomposition}
\label{Appen:A}
The mathematical tool utilized in this study is modal decomposition using spherical harmonics. We know that any given function can be expressed in terms of an orthonormal basis. Spherical harmonics form a complete set of one such orthonormal basis, used to express a function defined on the surface of a sphere. They serve as solutions to the angular part of Laplace’s equation in the spherical polar coordinate system. The mathematical expression for a spherical harmonic of degree $l$ and order $m$ is given by
 \begin{equation} 
 Y^m_l =
 \sqrt{\frac{2l +1}{4\pi}
 \frac{(l-m)!}
 {(l +m)!}}\, P^m_l (cos\,\theta)\, e^{im\phi}
 \label{eq:sphhar}
 \end{equation}
\noindent where $P^m_l (cos\, \theta)$ is referred to as the associated Legendre polynomial of degree $l$ and order $m$ while the symbols $\theta$ and $\phi$ denote colatitude and longitude, respectively. For a given $l$, $m$ varies from $-l$ to $+l$. We generally visualize spherical harmonics in terms of their nodal lines. A nodal line corresponds to the polarity inversion line connecting only those points on the sphere where either of the real or imaginary parts of the spherical harmonics, i.e., $Re[Y^m_l]$ or $Im[Y^m_l]$, becomes zero. The orthonormality condition is given by
 \begin{equation}
 \int_{\phi=0}^{2\pi}d\phi \int_{\theta=0}^{\pi}sin\, \theta \,d\theta \,Y^m_l(\theta, \phi)\, Y^{m'*}_{l'}(\theta, \phi)=\delta_{ll'}\,\delta_{mm'}
 \label{eq:ortho}
 \end{equation}

\noindent with $Y^{m'*}_{l'}$ being the complex conjugate of $Y^m_l$.

\begin{figure}[htbp]
  \centering
  \begin{minipage}{0.8\textwidth}
    \includegraphics[width=\linewidth]{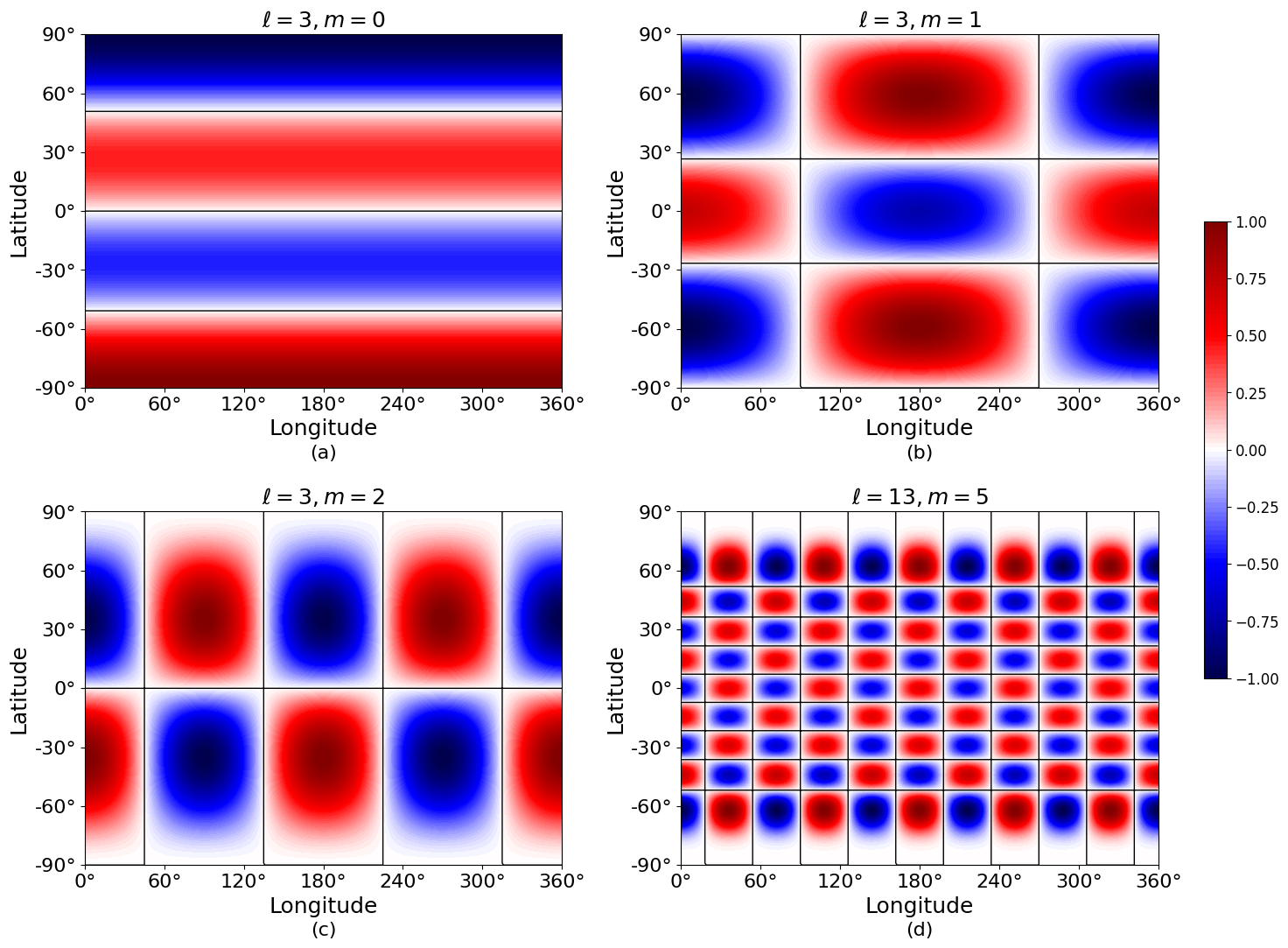}
\caption{Representation of $Y_l^m(\theta, \phi)$ for $l=3,\,m=0,1,2$ and $l=13,\,m=5$ are shown on a 2D Cartesian plane with nodal lines denoted in black. The color bar is to depict the sign of $Y_l^m(\theta, \phi)$.}
    \label{fig:sph}
    \end{minipage}
    \end{figure}

Figures \ref{fig:sph}(a)-(c) represent the spherical harmonics with degree $l = 3$ and orders $m=0,1,2$ distributed on a 2D Cartesian plane representing the surface of a sphere having $l-|m|$ latitudinal (horizontal) and $2|m|$ longitudinal (vertical) nodal lines. Figure \ref{fig:sph}(d) associated with $l = 13$ and $m = 5$ demonstrates how the surface resolution changes with different degree and order values. It is apparent that with an increasing degree ($l$), the number of nodal lines increases, which ultimately helps to resolve much finer details on the surface distribution. Mathematically, all possible degrees of spherical harmonics (i.e. $l = 0$ to $\infty$) are required to express any function in the spectral space. However, in general, one can use a finite value of the maximum degree ($l$) depending on the required resolution associated with the physical problem under consideration.

\section{Computational Model for the Solar Coronal Magnetic Field}
\label{Appen:B}
Modal decomposition is not only important for the photospheric magnetic field but also plays a vital role in simulating the coronal magnetic field distribution. The PFSSE model of the solar corona, proposed by \citet{1969SoPh....9..131A} and \citet{1969SoPh....6..442S}, is newly developed and used here for coronal magnetic field extrapolation using observed solar magnetogram data. Due to the low plasma $\beta$ ($< 1$), the model assumes a current-free corona in the space between the photosphere and an imaginary spherical surface typically located at $2.5 \, R_\odot$, known as the source surface ($R_{ss}$). Beyond $R_{ss}$, the magnetic field lines become radial as the solar wind starts to dominate over the dynamics of magnetic fields (plasma $\beta > 1$ ) and drags the open field lines into interplanetary space. This implies that the tangential components of the field are forced to become zero here, whereas the radial component has to be equal to the observed field distribution at the inner surface (photosphere). Combining the current-free assumption of PFSSE ($\nabla \times \textbf{B} = 0$) and solenoidal condition of magnetic field ($\nabla . \textbf{B}=0$), we get the Laplace equation such that 
 \begin{equation}
 \nabla^2\Psi = 0
 \label{eq:laplace}
 \end{equation}
 where 
 $\textbf{B}=-\nabla\Psi$ and $\Psi$ is the scalar potential. The mathematical implementation of boundary conditions suggests that at the inner boundary, $B_r (R_\odot, \theta, \phi) = B_r ^{obs}(\theta, \phi), \,\,\, \,\,\,i.e.- \frac{\partial \Psi}{\partial r}= B_r ^{obs}(\theta, \phi)$.
And, at the outer boundary, $B_\theta(R_{ss}, \theta, \phi) = 0, \,\,\,\,B_\phi(R_{ss}, \theta, \phi) = 0 $ \citep{Arden_2016}.
This PFSSE model can be developed using different numerical techniques such as spherical harmonics decomposition and the finite difference method. In this study, we solve Equation (\ref{eq:laplace}) using spherical harmonics to derive the field distribution in the global corona. The scalar potential is expressed as 
\begin{equation}
\Psi(r,\theta,\phi) = \sum_{l=1} ^{\infty} \sum_{m=-l} ^{l} C_{l}^{m}\, R_0\,\frac{(\frac{R_0}{r})^{l+1}[1-(\frac{r}{R_{ss}})^{2l+1}]}{l+1+l(\frac{R_0}{R_{ss}})^{2l+1}} Y_l ^m (\theta, \phi)
\label{eq:sclpot}
\end{equation}
where $R_0 = R_\odot$, $R_{ss} = 2.5$ $R_\odot$. Expressions of the extrapolated magnetic field components inside the coronal region are given by the following expressions \citep{1992ApJ...392..310W} for the radial component:
\begin{equation}
B_r(r, \theta, \phi) = - \frac{\partial \Psi}{\partial r} = \sum_{l=1} ^{\infty} \sum_{m=-l} ^{l} C_{l}^{m}\, (\frac{R_0}{r})^{l+2} \,[\frac{l+1+l(\frac{r}{R_{ss}})^{2l+1}}{l+1+l(\frac{R_0}{R_{ss}})^{2l+1}}]\,\, Y_l^m (\theta, \phi)
\label{eq:Br}
\end{equation}
and tangential components:
\begin{equation}
B_\theta(r, \theta, \phi) = - \frac{1}{r}\frac{\partial \Psi}{\partial \theta} = -\sum_{l=1} ^{\infty} \sum_{m=-l} ^{l} C_{l}^{m}\, (\frac{R_0}{r})^{l+2} \,[\frac{1-(\frac{r}{R_{ss}})^{2l+1}}{l+1+l(\frac{R_0}{R_{ss}})^{2l+1}}] \,\,\frac{\partial Y_l ^m (\theta, \phi)}{\partial \theta}
\label{eq:Btheta}
\end{equation}
and, \begin{equation}
B_\phi(r, \theta, \phi) = - \frac{1}{r\, sin \,\theta}\frac{\partial \Psi}{\partial \phi} = -\sum_{l=1} ^{\infty} \sum_{m=-l} ^{l} i\,m\,\,C_{l}^{m}\, (\frac{R_0}{r})^{l+2} \,[\frac{1-(\frac{r}{R_{ss}})^{2l+1}}{l+1+l(\frac{R_0}{R_{ss}})^{2l+1}}] \,\,\frac{Y_l ^m (\theta, \phi)}{sin \,\theta}
\label{eq:Bphi}
\end{equation}

\section{The maximum range of harmonic degree}
\label{Appen:C}
\begin{figure}
\begin{minipage}[t]{0.32\textwidth}
  \centering
  \includegraphics[width=\linewidth]{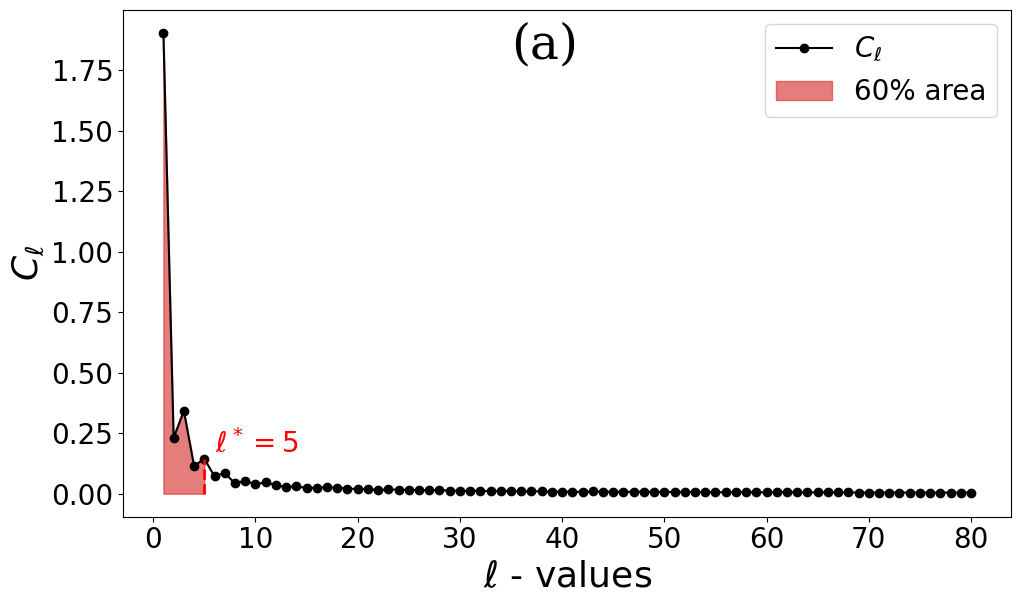}
\end{minipage}
\hfill
\begin{minipage}[t]{0.32\textwidth}
  \centering
  \includegraphics[width=\linewidth]{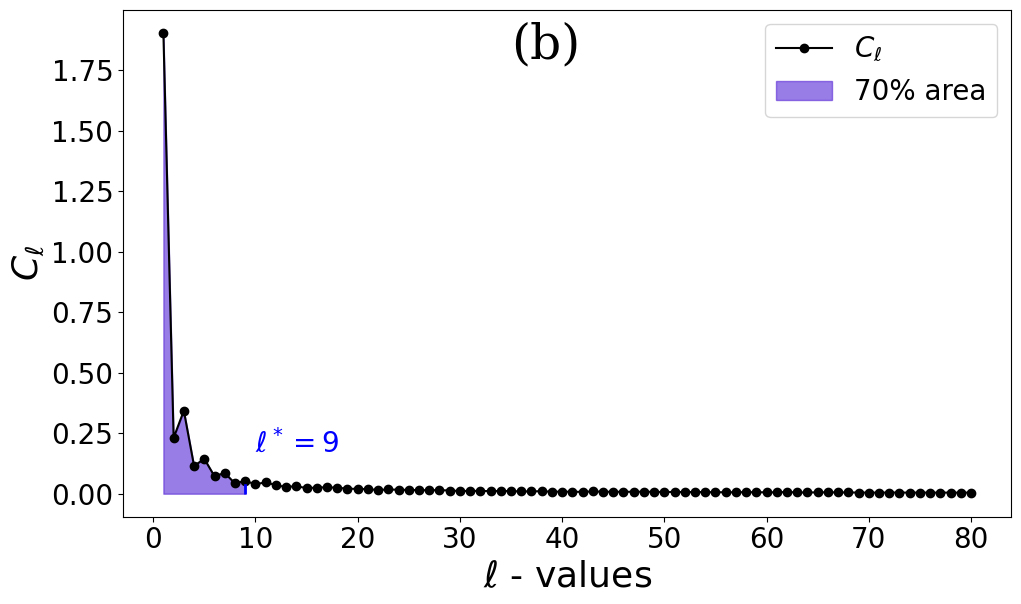}
\end{minipage}
\hfill
\begin{minipage}[t]{0.32\textwidth}
  \centering
  \includegraphics[width=\linewidth]{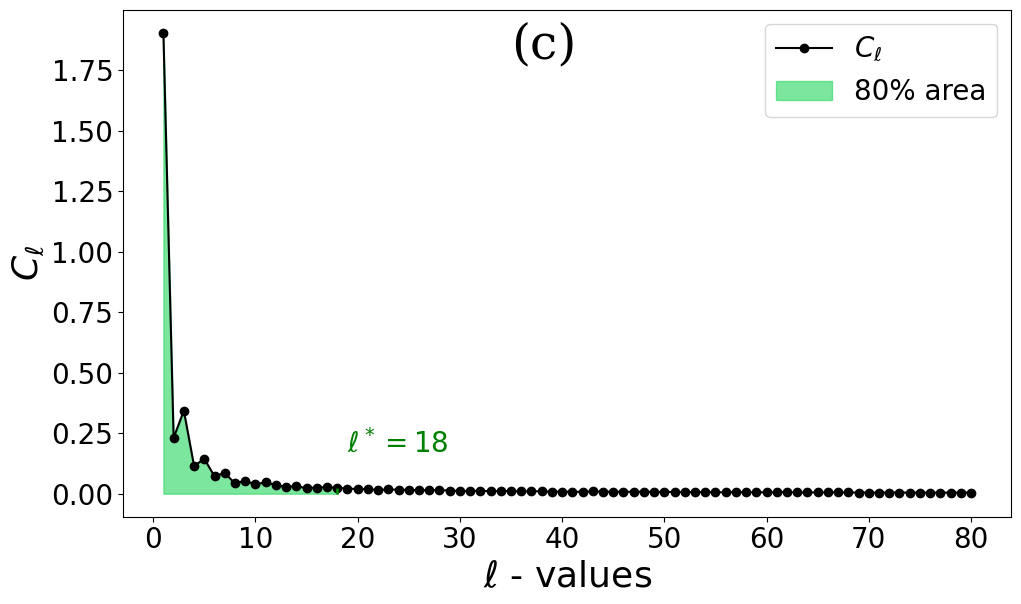}
\end{minipage}
\caption{Effective mode contribution spectrum for CR 2079, where the shaded regions represent (a) $60\%$, (b) $70\%$, and (c) $80\%$ of the total area under the curve, respectively. The limiting values of degree $l$ for each contribution are specified in the plots where (a) $l^* = 5$, (b) $l^* = 9$, and (c) $l^* = 18$  correspond to $60\%$, $70\%$, and $80\%$ contributions, respectively.}
\label{fig:678}
\end{figure}

\begin{figure}
  \centering
  \begin{minipage}{0.9\textwidth}
    \includegraphics[width=\linewidth]{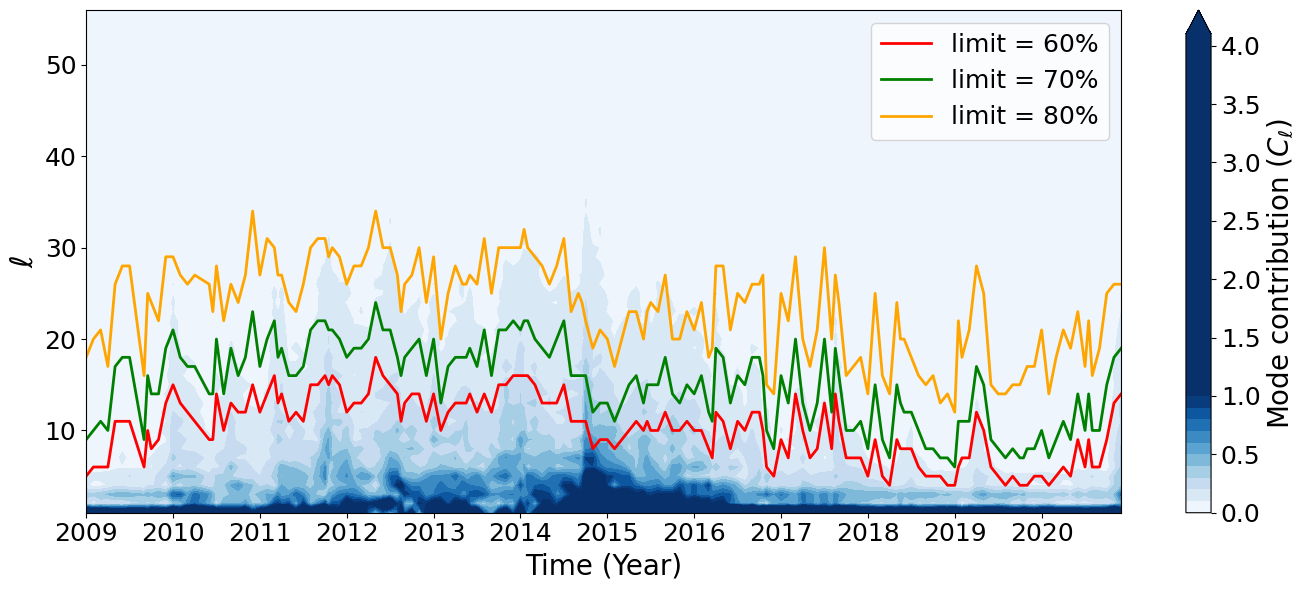}
    \caption{Effective mode contribution spectrum superimposed with the limiting degree ($l^*$) curves associated with $60 \%$ (red), $70\%$ (green), and $80\%$ (orange) contributions during solar cycle 24. The color bar is saturated at a maximum mode strength of $C_l = 1.0$.}
    \label{fig:con678}
    \end{minipage}
\end{figure}

The analysis at the end of the third paragraph of subsection \ref{subsec:result_mode_analysis} regarding the evaluation of $l^*$ is presented here in detail. To find a quantitative limit of $l$, up to which we should consider the modal contributions, we further analyze the contributions for $60\%$, $70\%$, and $80\%$ of the total spectral coefficient (area under the curve of the spectral-mode contributions) for each Carrington map. One such analysis is demonstrated in Figure \ref{fig:678} for the minima of Cycle 24 during 2009 January, corresponding to CR 2079. Here, $80\%$ of the contribution of the effective mode is achieved by $l^* = 18$. Extending this analysis across solar cycle 24, Figure \ref{fig:con678} shows the resulting $l^*$ values for each threshold, overlaid on the modal contribution contours. The red, green and orange curves imply $l^*$ corresponding to $60\%$, $70\%$, and $80\%$, respectively.

\section{Effect of a single sunspot}

\label{Appen:D}
\begin{figure}
  \centering
  \begin{minipage}{0.8\textwidth}
  \centering
    \includegraphics[width=\linewidth]{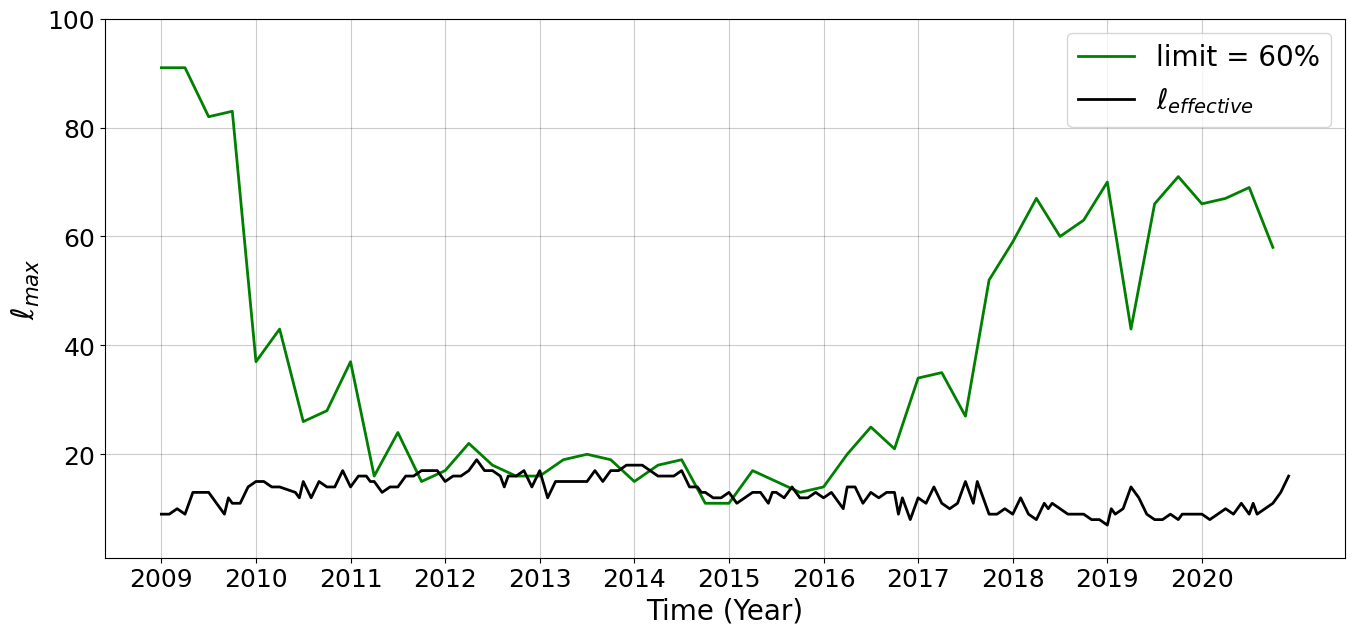}
    \caption{Variation of $l_\mathrm{max}$ throughout solar cycle 24 related to the $60\%$ coverage of observed photospheric flux (green curve) plotted along with the photospheric $l_\mathrm{effective}$ (black curve).}
    \label{fig:fluxphoto}
    \end{minipage}
\end{figure}

We also analyze photospheric magnetic flux variations over time to address the same question of the required maximum resolution. For this, a total of $48$ Carrington maps (four for each year) are taken into consideration. We inspect the maximum degree ($l_\mathrm{max}$) required to cover $60\%$ of the total unsigned photospheric flux computed from the observed MDI and HMI magnetogram data. The green curve in Figure \ref{fig:fluxphoto} denotes the variation of $l_\mathrm{max}$ throughout solar cycle 24 for the $60\%$ coverage of the observed photospheric flux, while the black curve corresponds to the variation of the photospheric effective degree ($l_\mathrm{effective}$) correlated with the modal contribution. Although these two curves are comparable with each other during solar cycle 24 maxima (end of 2011 to beginning of 2016), they start diverging from each other as we move to either side of the maxima. The difference during the solar minima becomes significant with $l_\mathrm{effective} \approx 10$, but $l_\mathrm{max} > 60$ for $60\%$ flux coverage.

It suggests that multiple sunspots present during solar maxima on the photosphere collectively appear as large clusters of magnetic field corresponding to lower values of $l_\mathrm{max}$ capable of $60\%$ flux coverage. However, a high value of $l_\mathrm{max}$ during minima relates to the presence of one or two strong field regions, such as decaying sunspots or AR remnants, with a backdrop of an overall quiet Sun. If such strong magnetic field islands are omitted from the data through smoothing (based on the Gaussian filter), $l_\mathrm{max}$ reduces by a significant amount. Thus, while evaluating the effect of a single sunspot or a group of sunspots on the global field distribution through relative flux content, one must practice caution with the interpretation.

\end{document}